\documentclass[aps,prappl,reprint,groupedaddress]{revtex4-1}

\usepackage{graphicx}
\usepackage{dcolumn}
\usepackage{bm}
\usepackage{amsmath,amssymb}
\usepackage{comment}

\def\U#1{{%
\def\O{\mbox{O}}
\def\u{\mbox{u}}
\mathcode`\u=\mu
\mathcode`\O=\Omega
\mathrm{#1}}}

\def\ii{{\mathrm{i}}}
\def\dd{{\mathrm{d}}}
\def\sub#1{_{\scriptsize\mbox{#1}}}

\def\dquote#1{\textquotedblleft #1\textquotedblright}
\def\degree{\mbox{$^\circ$}}

\begin{document}


\title{Broadband control of group delay using the Brewster effect in metafilms}
\author{Yasuhiro Tamayama}
\email{tamayama@vos.nagaokaut.ac.jp}
\author{Hiromu Yamamoto}
\affiliation{Department of Electrical, Electronics and
Information Engineering, Nagaoka University of
Technology, 1603-1 Kamitomioka, Nagaoka, Niigata 940-2188, Japan}

\date{\today}

\begin{abstract}
We propose and verify a method for controlling the frequency dependence of the group delay of electromagnetic waves over a broad frequency range using the Brewster effect in single-layer metamaterials with finite thickness, here referred to as metafilms. When the metafilm's reflectance vanishes regardless of the incident frequency, the group delay can be large near its resonance frequency while maintaining the transmittance close to unity regardless of the incident frequency. Furthermore, when several reflectionless metafilms are stacked together, the total group delay should be given as the sum of the individual group delays. In this study, we realize reflectionless metafilms by arranging the meta-atoms so that the Brewster effect occurs regardless of the incident frequency. We evaluate in numerical simulations and experiments the frequency dependence of the transmittance and of the group delay of a three-layer metamaterial composed of reflectionless metafilms with different resonance frequencies, and find that the total transmittance and group delay of this metamaterial agree respectively with the product of the transmittances and the sum of the group delays of the constituent metafilms.

\end{abstract}


\maketitle

\section{Introduction}

The slow-group-velocity propagation~\cite{hau_99_nature} and storage~\cite{liu_01_nature,phillips_01_prl} of electromagnetic waves have been realized using a quantum interference effect known as electromagnetically induced transparency (EIT)~\cite{fleischhauer_05_rmp}. To date, extensive efforts have been devoted to mimic EIT in metamaterials for realizing slow-group-velocity propagation and other related phenomena over various frequency bands. These efforts have been based on mechanical and electrical circuit models of EIT~\cite{alzar_02_ajp}. To realize a metamaterial that mimics EIT, these classical models have revealed that the unit cell of the metamaterial should be composed of two resonators, one of low $Q$ that is excited by incident waves and coupled to the other of high $Q$ that is not excited directly by the incident waves~\cite{zhang_s_08_prl,tassin_09_prl,liu_n_09_nat_mater,lu_10_opex,tamayama_12_prb, rana_18_prappl,bagci_18_jap,xu_z_19_prb}. Metamaterials with an EIT-like transmission property have also been realized by arranging two coupled resonators to be excited directly by the incident waves~\cite{jin_x-r_12_jap,tamayama_14_prb,tamayama_15_prb, hokmabadi_15_sci_rep}. Storage of the electromagnetic waves has even been realized using metamaterials that mimic EIT~\cite{nakanishi_13_prb,nakanishi_18_apl}, as was demonstrated using the original EIT effect.

Although EIT-like metamaterials are playing an important role in controlling the group velocity (group delay) of electromagnetic waves, there are hints that they may not be the best solution. In EIT-like metamaterials, the large group delay occurs over a narrow frequency range and incident electromagnetic waves outside of this narrow frequency range are reflected. Therefore, controlling the frequency dependence of the group delay using EIT-like metamaterials over a broad frequency range may be difficult.

Recently, metasurfaces with multiple Lorentzian electric and magnetic resonances have enabled broadband control of the group delay~\cite{ginis_16_apl,tsilipakos_21_acsphoton}. In this study, we demonstrate broadband control of the group delay using the Brewster effect in a single-layer metamaterial with only an electric response. Hereafter, we refer to a single-layer metamaterial with finite thickness as a metafilm. So far, the Brewster effect has been investigated in various kinds of metamaterials and metasurfaces~\cite{tamayama_06_prb,alu_11_prl,dominguez_16_nat_comm,wang_c_19_apl,yin_s_19_opex,lavigne_21_opex,fan_h_21_prappl,zhang_21_prappl}. As a part of these studies, we showed in our previous papers~\cite{tamayama_15_ol,tamayama_16_ol} that reflection in a metafilm can be completely suppressed by arranging the meta-atoms so that the oscillation direction of the electric dipole (or magnetic dipole) induced in the meta-atoms coincides with the propagation direction of the reflected wave. This idea is based on the physical mechanism underlying the Brewster effect. Hereafter, we refer to a metafilm that is designed to ensure the Brewster effect occurs as a Brewster metafilm. When the oscillation direction of the electric dipole induced in the meta-atom is independent of the incident frequency, the reflectance of the metafilm vanishes regardless of the incident frequency. In this instance, the absolute value of the transmittance is unity regardless of the incident frequency, and the group delay becomes large near the resonance frequency of the Brewster metafilm if non-radiation losses of the constituent meta-atoms are sufficiently small~\cite{tamayama_16_ol}. This implies that broadband control of the group delay can be achieved by stacking Brewster metafilms with different resonance frequencies. In the following, we verify this idea through numerical analyses and microwave experiments. 

\section{Theory}

\begin{figure}[tb]
\begin{center}
\includegraphics[scale=1]{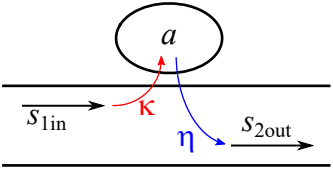}
\caption{Schematic of a resonator system coupled with input and output ports. To model the reflectionless property of the resonator, the radiated wave from the resonator is assumed to propagate only in the propagation direction of the transmitted wave.}
\label{fig:cmt}
\end{center}
\end{figure}

\begin{figure}[tb]
\begin{center}
\includegraphics[scale=0.65]{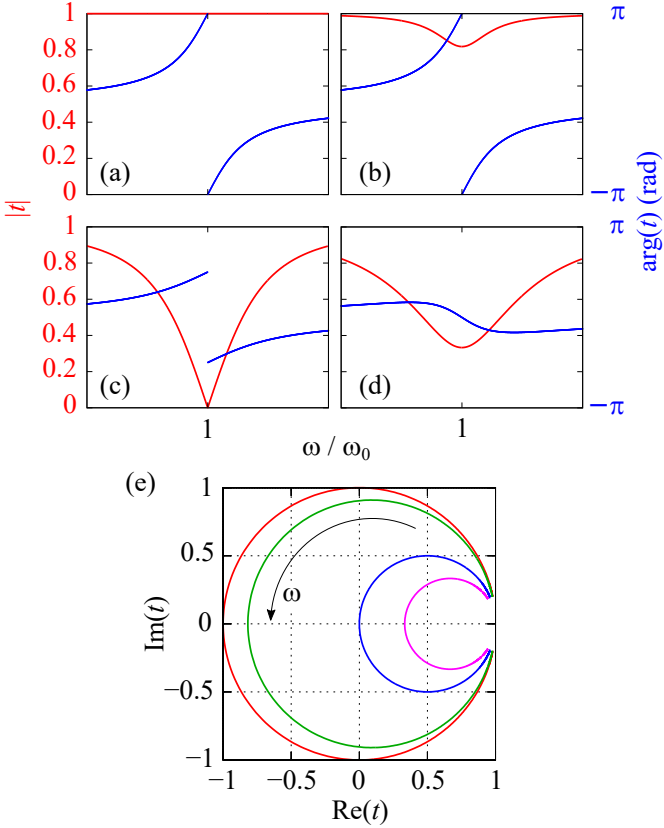}
\caption{Angular frequency dependence of the complex transmittance given by Eq.~\eqref{eq:30} for $\gamma\sub{nr}/\gamma\sub{r}$ of (a) 0, (b) 0.1, (c) 1, and (d) 2. (e) Locus of the complex transmittance as a function of angular frequency in the complex plane. The curves correspond to $\gamma\sub{nr}/\gamma\sub{r}=0$, 0.1, 1, and 2 in order from large to small radius. The angular frequency range in (a)--(d) is $\omega_0 - 4\gamma\sub{r} < \omega < \omega_0 + 4\gamma\sub{r}$ and that in (e) is $\omega_0 - 10\gamma\sub{r} < \omega < \omega_0 + 10\gamma\sub{r}$. $\omega_0/\gamma\sub{r} = 10$ is assumed in the calculation.}
\label{fig:theory_trans}
\end{center}
\end{figure}

First, we analyze the transmission property of a reflectionless metafilm using a model (Fig.~\ref{fig:cmt}) in which we assume there exist a propagation mode and a resonance mode. The resonance mode corresponds to a resonant metafilm. To model the reflectionless property of the metafilm, the radiated wave from the resonator is assumed to propagate only in the direction of propagation of the transmitted wave. Based on the coupled mode theory~\cite{suh_04_ieee,ruan_10_jpc,ruan_10_prl}, we obtain the governing equations,
\begin{align}
\frac{\dd a}{\dd \tau}
&=(-\ii \omega_0 - \gamma\sub{r} - \gamma\sub{nr} )a + \kappa s\sub{1in} ,\\
s\sub{2out} 
&=s\sub{1in} + \eta a ,
\end{align}
where $s\sub{1in}$ ($s\sub{2out}$) denote the amplitude of the incident (transmitted) wave, $a$ the amplitude of the resonance mode, $\omega_0$ the resonance angular frequency of the resonator, $\gamma\sub{r}$ ($\gamma\sub{nr}$) the radiation loss (non-radiation loss) of the resonator, $\kappa$ ($\eta$) the coupling between the incident (transmitted) wave and the resonance mode, and $\tau$ time. Using the relation $\kappa \eta = -2\gamma\sub{r}$, which is derived from energy conservation and time-reversal symmetry, the amplitude transmittance for a continuous wave with angular frequency of $\omega$ is calculated to be
\begin{equation}
t = \frac{s\sub{2out}}{s\sub{1in}} = \frac{-\ii (\omega - \omega_0) - \gamma\sub{r} 
+ \gamma\sub{nr}}{-\ii (\omega - \omega_0) + \gamma\sub{r} + \gamma\sub{nr}} .
\label{eq:30}
\end{equation}
Figures \ref{fig:theory_trans}(a)--\ref{fig:theory_trans}(d) show the angular frequency dependence of this complex transmittance for various values of $\gamma\sub{nr}/ \gamma\sub{r}$. The absolute value of the amplitude transmittance $|t|$ is unity regardless of the angular frequency for $\gamma\sub{nr}/ \gamma\sub{r} =0$, and $|t|$ at $\omega = \omega_0$ decreases with increasing $\gamma\sub{nr}/ \gamma\sub{r}$ for $\gamma\sub{nr}/ \gamma\sub{r} < 1$. At $\omega = \omega_0$, $|t|$ vanishes; that is, perfect absorption occurs at $\omega = \omega_0$ for $\gamma\sub{nr}/ \gamma\sub{r} =1$, which is referred to as critical coupling~\cite{ruan_10_prl}. For $\gamma\sub{nr}/ \gamma\sub{r} > 1$, $|t|$ at $\omega = \omega _0$ increases with increasing $\gamma\sub{nr}/ \gamma\sub{r}$. The group delay, which is given by $\dd [\arg{(t)}] / \dd \omega$, has a large positive value near $\omega = \omega_0$ for $\gamma\sub{nr}/ \gamma\sub{r} < 1$ and becomes negative near $\omega = \omega_0$ for $\gamma\sub{nr}/ \gamma\sub{r} > 1$. The angular frequency dependence of the complex transmittance is quite different depending on whether $\gamma\sub{nr}/ \gamma\sub{r} < 1$ or $\gamma\sub{nr}/ \gamma\sub{r} > 1$. The reason for this difference is understood from the locus of $t$ in the complex plane [Fig.~\ref{fig:theory_trans}(e)]. The position of $t$ moves in a counterclockwise direction in the complex plane as the angular frequency increases. The size of the locus of $t$ decreases with increasing $\gamma\sub{nr}/ \gamma\sub{r}$. This figure implies that the difference is caused by whether the point at the intersection of the locus of $t$ with the real axis is positive or negative. (Note that the intersection corresponds to $\omega = \omega _0$.) The group delay near $\omega = \omega_0$ is found to be positive (negative) when the intersection is on the negative (positive) real axis because the point of $t$ moves in a counterclockwise direction with increasing angular frequency. The complex transmittance at $\omega = \omega_0$ is also found to vary only between $-1$ and $1$ as $\gamma\sub{nr}/ \gamma\sub{r}$ increases.

From the above analysis, the transmittance and group delay of a metafilm near the resonance frequency are found to be controlled through varying $\gamma\sub{nr} / \gamma\sub{r}$ if the reflection is completely suppressed regardless of the incident frequency. When such reflectionless metafilms are stacked, the total complex transmittance of the stacked metafilms should be given by the product of the complex transmittances of the constituent metafilms because of the reflectionless property of the metafilms. This implies that a multilayer metamaterial with a specified complex transmission spectrum may be designed by controlling $\omega_0$ and $\gamma\sub{nr}/\gamma\sub{r}$ of each metafilm layer. In particular, for group-delay control over a broad frequency range, we need only to stack reflectionless metafilms with different resonance frequencies and $\gamma\sub{nr} / \gamma\sub{r} \ll 1$. 

\section{Simulation and experiment}

\begin{figure}[tb]
\begin{center}
\includegraphics[scale=0.9]{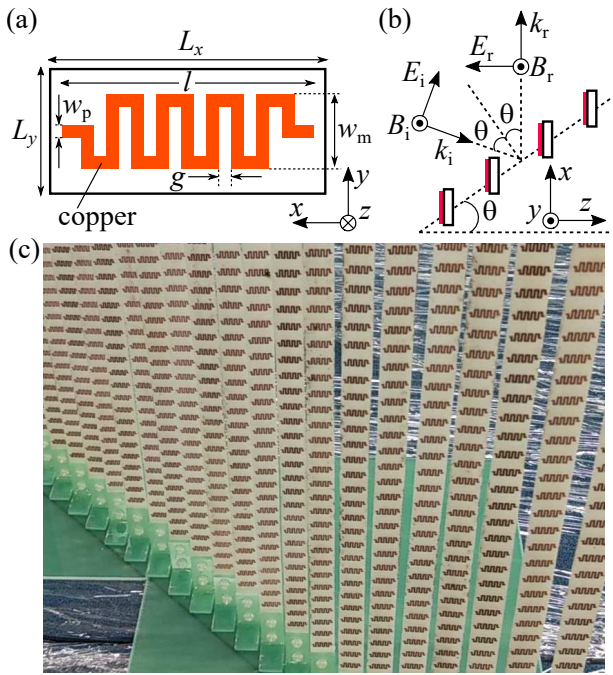}
\caption{(a) Schematic of the structure of the meta-atom used in studying broadband control of group delay. The parameter settings used are: $w\sub{p}=0.5~\U{mm}$, $g=0.5~\U{mm}$, $l=9.5~\U{mm}$, $L_x = 11.5~\U{mm}$, $L_y = 7.5~\U{mm}$. The range of values for $w\sub{m}$ is given in the main text. The thickness of the substrate is 0.8~mm. (b) Relationship between the electromagnetic fields and the distribution of the meta-atoms needed to produce the Brewster effect in a metafilm. The subscripts \dquote{i} and \dquote{r} indicate the incident and reflected fields, respectively. (c) Photograph of one of the fabricated Brewster metafilms. }
\label{fig:setup}
\end{center}
\end{figure}

From numerical and experimental work, we verified the above theory for broadband control of the group delay using Brewster metafilms as reflectionless metafilms. We use meander-line resonators [Fig.~\ref{fig:setup}(a)] as the meta-atoms that constitute the Brewster metafilms because such resonators have a relatively high radiation loss, which is advantageous in decreasing $\gamma\sub{nr} / \gamma\sub{r}$. When an $x$-polarized electromagnetic wave is incident on the meander-line resonator, an electric dipole is induced in the $x$ direction. Therefore, a Brewster metafilm may be fabricated by arranging the meander-line resonators so that the propagation direction of the reflected wave coincides with the $x$ direction [Fig.~\ref{fig:setup}(b)]. To minimize $\gamma\sub{nr} / \gamma\sub{r}$ for a given $\gamma\sub{nr}$, that is, to maximize $\gamma\sub{r}$, the propagation direction of both the incident and transmitted wave should coincide with the direction for which the radiation pattern of the electric dipole induced in the resonator is maximal~\cite{tamayama_19_josab}. Therefore, in the following, we set $\theta=45\degree$.

\begin{figure}[tb]
\begin{center}
\includegraphics[scale=1]{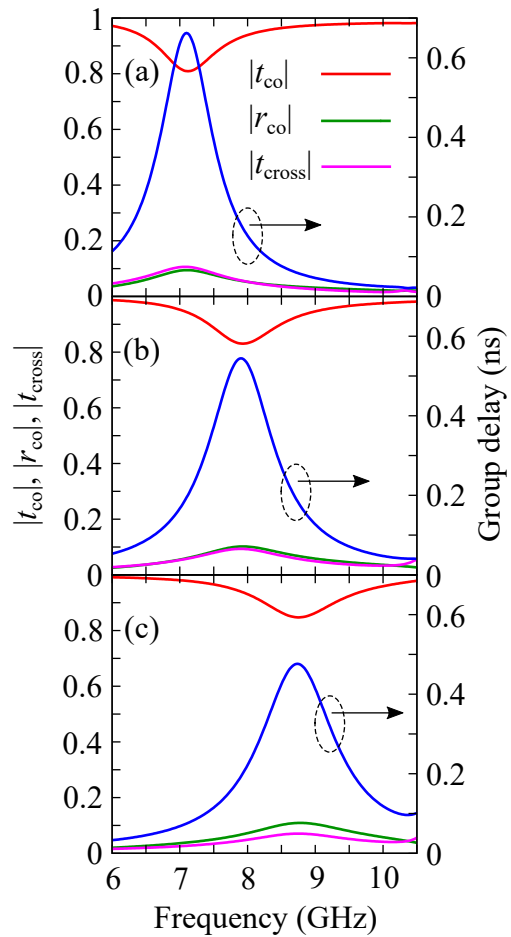}
\caption{Transmission and reflection spectra of Brewster metafilms obtained from numerical calculations with  $w\sub{m}$ of (a) $3.0~\U{mm}$, (b) $2.5~\U{mm}$, and (c) $2.0~\U{mm}$. The relative permittivity of the substrate is assumed to be $4.5(1+\ii 0.03)$ for this calculation.}
\label{fig:sim_single}
\end{center}
\end{figure}

\begin{figure}[tb]
\begin{center}
\includegraphics[scale=1]{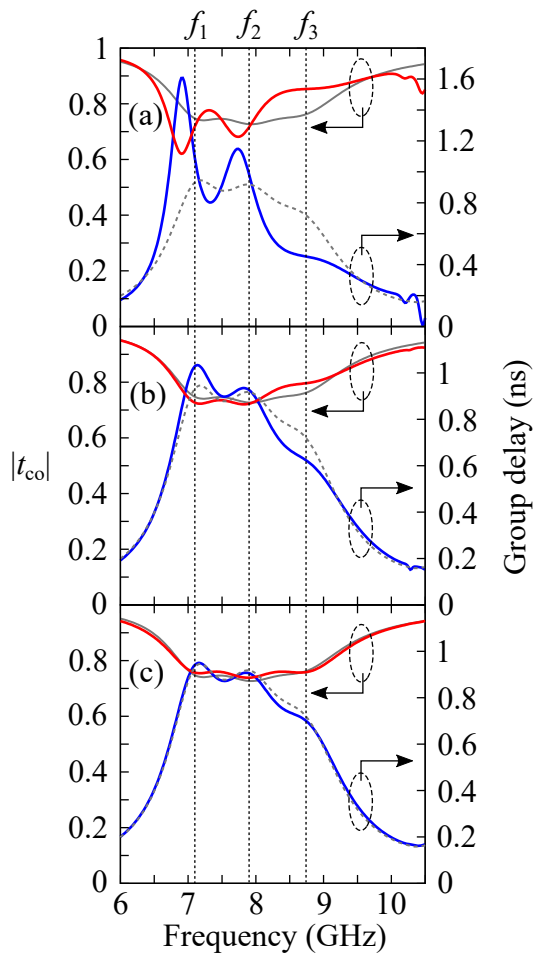}
\caption{Transmission spectra of the three-layer metamaterial obtained from numerical calculations with (a) $d_z= 10~\U{mm}$, (b) 20~mm, and (c) 40~mm. The gray solid and dashed curves are the respective plots of the product of the individual transmittances $|t\sub{co}|$ and the sum of the individual group delays of the three constituent Brewster metafilms. The resonance frequencies of the Brewster metafilms with $w\sub{m}= 3.0~\U{mm}$, $2.5~\U{mm}$, and $2.0~\U{mm}$ are $f_1= 7.10~\U{GHz}$, $f_2 = 7.90~\U{GHz}$, and $f_3 = 8.74~\U{GHz}$, respectively. The relative permittivity of the substrate is assumed to be $4.5(1+\ii 0.03)$ for this calculation. }
\label{fig:sim_3layers}
\end{center}
\end{figure}

Using commercial finite-element software COMSOL Multiphysics, we analyzed the reflection and transmission properties of the designed metafilms. In this numerical analysis, we assumed that the meander-line resonator was made of a perfect electric conductor with vanishing thickness, and that the substrate was FR-4 with a relative permittivity of $4.5(1+\ii 0.03)$. We calculated the co-polarized amplitude reflectance $r\sub{co}$, co-polarized amplitude transmittance $t\sub{co}$, and cross-polarized amplitude transmittance $t\sub{cross}$ for metafilms with $w\sub{m}= 3.0~\U{mm}$, 2.5~mm, and 2.0~mm (Fig.~\ref{fig:sim_single}). For all instances, $|r\sub{co}|$ and $|t\sub{cross}|$ in the calculated frequency range are less than about 0.1, which implies that the reflection and the polarization conversion is small. From this result, Brewster metafilms may be fabricated using meander-line resonators as meta-atoms. The group delay, which is given by $\dd [\arg{(t\sub{co})}] / \dd \omega$, peaks at resonance frequencies 7.10~GHz, 7.90~GHz, and 8.74~GHz when $w\sub{m}= 3.0~\U{mm}$, 2.5~mm, and 2.0~mm, respectively. (Throughout, the value of the group delay is defined as the difference between the group delays from the transmitter to the receiver of the system with and without the metafilm(s).) The value of $|t\sub{co}|$ at resonance decreases from unity because of the dielectric loss of the substrate (to be discussed later).
Next, we analyzed the transmission properties of a three-layer metamaterial composed of all three metafilms set with $w\sub{m}= 3.0~\U{mm}$, 2.5~mm, and 2.0~mm. Because of the reflectionless property of the metafilms, $|t\sub{co}|$ and the group delay of the three-layer metamaterial should agree with the product of the individual transmittances $|t\sub{co}|$ and the sum of the individual group delays of the constituent metafilms, respectively. Figure \ref{fig:sim_3layers} shows plots of the transmittance $|t\sub{co}|$ and the group delay for the three-layer metamaterial obtained from numerical calculations with the interlayer distance between adjacent metafilms in the $z$ direction of $d_z = 10~\U{mm}$, 20~mm, 40~mm. For $d_z = 10~\U{mm}$, the difference between the transmittance $|t\sub{co}|$ (group delay) of the three-layer metamaterial and the product of transmittances $|t\sub{co}|$ (sum of the group delays) of the constituent metafilms is large. The difference becomes quite small when $d_z$ exceeds 20~mm. For $d_z = 40~\U{mm}$, the transmittance $|t\sub{co}|$ (group delay) of the three-layer metamaterial almost agrees with the product of transmittances $|t\sub{co}|$ (sum of the group delays) of the constituent metafilms.
This result implies that the difference arises from the interlayer coupling between the constituent metafilms. Therefore, when the constituent metafilms are not mutually coupled, the transmittance (group delay) of a multilayer metamaterial composed of stacked Brewster metafilms may agree with the product of the transmittances (sum of the group delays) of the constituent metafilms. 

\begin{figure}[tb]
\begin{center}
\includegraphics[scale=1]{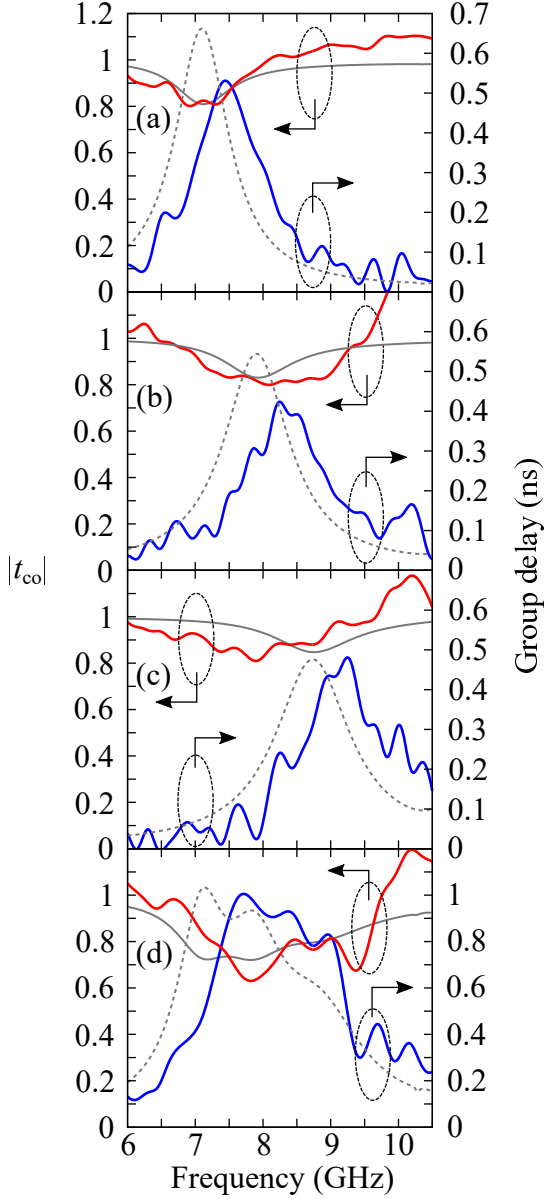}
\caption{Measured frequency dependences of transmittance $|t\sub{co}|$ and the group delay of the Brewster metafilms, individually with (a) $w\sub{m}= 3.0~\U{mm}$, (b) $2.5~\U{mm}$, and (c) $2.0~\U{mm}$, and (d) those for the three-layer metamaterial with $d_z = 20~\U{mm}$. The value of $|t\sub{co}|$ is defined as the ratio of the transmittances with and without the metafilm(s) between the transmitting and receiving antennas. Because the size of the metafilm(s) and the beamwidth of the electromagnetic wave are finite, $|t\sub{co}|$ exceeds unity in some frequency regions. The gray solid and dashed curves are plots of $|t\sub{co}|$ and the group delay presented in Figs.~\ref{fig:sim_single} and \ref{fig:sim_3layers}(b) for reference.}
\label{fig:trans}
\end{center}
\end{figure}

Following the numerical verification of the theory, we fabricated the designed metafilms to verify the theory through experiments. The meta-atoms were fabricated using printed circuit boards that consisted of a 35-$\U{um}$-thick copper film on a 0.8 mm thick FR-4 substrate. A photograph of one of the fabricated metafilm is shown in Fig.~\ref{fig:setup}(c). Although the meta-atoms were fixed to the stage in this study to simplify our proof-of-principle experiment, in practical applications, the meta-atoms may be fixed to a frame.

Figures \ref{fig:trans}(a)--\ref{fig:trans}(c) show the measured frequency dependences of $|t\sub{co}|$ and group delay for the three metafilms, individually with $w\sub{m}=3.0~\U{mm}$, 2.5~mm, and 2.0~mm; their respective group delays peak at resonance frequencies 7.4~GHz, 8.4~GHz, and 9.1~GHz. The transmittances at these resonance frequencies decrease from unity through dielectric losses in the substrate as observed from the numerical simulation. The measured and simulation results almost agree, implying that Brewster metafilms have been realized in these experiments using meander-line resonators as meta-atoms. 

Next, we measured the transmission properties of a three-layer metamaterial composed of the three metafilms with $w\sub{m}=3.0~\U{mm}$, 2.5~mm, and 2.0~mm. In the results of the numerical simulation, the transmittance (group delay) of the three-layer metamaterial almost agrees with the product of the individual transmittances (sum of the individual group delays) of the constituent metafilms when the interlayer distance $d_z$ is larger than 20~mm. Therefore, in the experiment, $d_z$ was set to 20~mm. The measured frequency dependences of the transmittance and group delay of the three-layer metamaterial [Fig.~\ref{fig:trans}(d)] show that the transmittance does not exhibit a steep variation and the group delay becomes large in value over a broader frequency range than instances using a single metafilm because of the difference in the resonance frequencies of the constituent metafilms. Also, the measured transmittance and group delay roughly agree with the numerical results. This theory for broadband control of group delay using multilayer Brewster metafilms is thus verified in both numerical simulations and experiments.

\begin{figure}[tb]
\begin{center}
\includegraphics[scale=1]{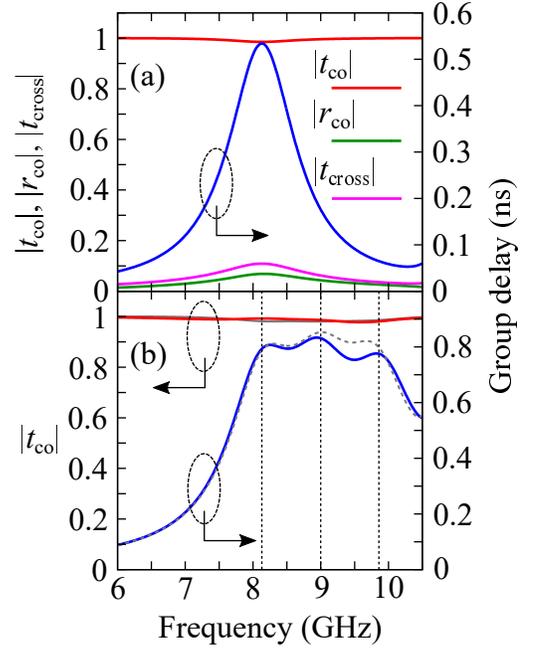}
\caption{(a) Transmission and reflection spectra of the Brewster metafilms with $w\sub{m}= 3.0~\U{mm}$ and (b) transmission spectrum of the three-layer metamaterial with $d_z = 40~\U{mm}$ obtained from numerical calculations. The gray solid and dashed curves are the product of the transmittances $|t\sub{co}|$ and the sum of the group delays of the Brewster metafilms with $w\sub{m}= 3.0~\U{mm}$, $2.5~\U{mm}$, and $2.0~\U{mm}$, respectively. The vertical dashed lines indicate the respective resonance frequencies, 8.13~GHz, 9.00~GHz, and 9.86~GHz, of these Brewster metafilms. The relative permittivity of the substrate is assumed to be $2.2(1+\ii 0.0009)$ for this calculation. }
\label{fig:lowloss}
\end{center}
\end{figure}

In fabricating the Brewster metafilms for this study, we used printed circuit boards with a FR-4 substrate because they are widely available at low cost. Although they are satisfactory for our proof-of-principle experiment, we note that the transmittances of the designed Brewster metafilms decrease from unity near the resonance frequency because of non-radiation losses in the metafilms. According to the above theory, the transmittance at the resonance frequency increases with decreasing $\gamma\sub{nr}/\gamma\sub{r}$ for $\gamma\sub{nr}/\gamma\sub{r}<1$. To estimate the attainable transmittance at the resonance frequency under realistic conditions, we analyze numerically the electromagnetic responses of the Brewster metafilms made of commercially available low-loss laminates, specifically, Rogers RT/duroid 5880. 
For the analysis, the relative permittivity of the substrate of the laminate was assumed to be $2.2(1+\ii 0.0009)$. Figure \ref{fig:lowloss}(a) shows the numerically calculated transmission and reflection properties of the metafilm with $w\sub{m}=3.0~\U{mm}$. Both $|r\sub{co}|$ and $|t\sub{cross}|$ are low in value in the calculated frequency range, as for the FR-4 substrate; moreover, $|t\sub{co}|$ at the resonance frequency (8.13~GHz) is higher than 0.98, which is higher than for the FR-4 substrate. From this result, the decrease in $|t\sub{co}|$ at the resonance frequency for the printed circuit board with the FR-4 substrate is found to occur because of dielectric losses in the substrate and, for the commercially available printed circuit board with the low-loss substrate, $|t\sub{co}|$ is found to be near unity at the resonance frequency. Figure \ref{fig:lowloss}(b) shows the transmission property of a three-layer metamaterial composed of metafilms with $w\sub{m}=3.0~\U{mm}$, 2.5~mm, and 2.0~mm. Because of lower dielectric losses in the substrate, the group delay becomes large over a broad frequency range while $|t\sub{co}|$ remains closer to unity than for the FR-4 substrate. The transmittance (group delay) of the three-layer metamaterial also agrees in this instance with the product of transmittances (sum of the group delays) of the constituent metafilms.

\section{Conclusion}
From numerical simulations and experiments, we verified a theory concerning the control of the group delay over a broad frequency range using Brewster metafilms. The transmittance of the Brewster metafilm with $\gamma\sub{nr}/\gamma\sub{r} \ll 1$ is almost unity regardless of frequency, and its group delay becomes large near the resonance frequency. Therefore, by stacking Brewster metafilms having different resonance frequencies, the frequency dependence of the group delay may be controlled over a broad frequency range while maintaining the transmittance close to unity. To verify this theory, we investigated a three-layer metamaterial composed of three Brewster metafilms with different resonance frequencies. When the near-field coupling between the constituent metafilms is sufficiently weak, the transmittance and the group delay of the three-layer metamaterial agree with the product of transmittances and the sum of the group delays of the constituent metafilms, respectively. We demonstrated that the frequency bandwidth for large group delay can be broadened by stacking Brewster metafilms as a proof-of-principle experiment of broadband control of the group delay. Furthermore, this theory also applies to other kinds of control of the group delay. For example, a multilayer metamaterial with positive (negative) group delay dispersion may be realized by stacking Brewster metafilms with different resonance frequencies so that the group delay in the higher frequency range becomes larger (smaller). Also, when we divide a dispersionless metamaterial composed of multiple Brewster metafilms into two metamaterials and place one behind the transmitter and the other in front of the receiver, the envelope of the electromagnetic wave generated from the transmitter is distorted intentionally in the region between the two metamaterials but this distortion is compensated at the receiver, in a manner reminiscent of an analog encryption--decryption system.

As described in our previous paper~\cite{tamayama_16_ol}, the macroscopic properties of Brewster metafilms are similar to those of Huygens metasurfaces with spectrally overlapping electric and magnetic resonances~\cite{decker_15_aom,chong15,asadchy15,balmakou15,
cuesta_18_ieee,londono_18_prappl,fathnan_20_aom,risco_21_prb}. In designing Huygens metasurfaces, both the electric and magnetic responses in the metasurfaces need to be controlled. In contrast, Brewster metafilms can be designed based on either their electric or magnetic responses and may be useful in realizing anisotropic reflectionless metamaterials. Although the fabrication of Brewster metafilms is more difficult than for Huygens metasurfaces, terahertz (optical) Brewster metafilms are possible to be fabricated using microelectromechanical systems~\cite{tao_09_prl} (3D laser lithography~\cite{gansel09} and 3D nanoprinting techniques~\cite{jung_w_21_nat}). There are many studies on controlling electromagnetic waves using Huygens metasurfaces~\cite{chen_18_np}, which shows the importance of reflectionless metamaterials in controlling electromagnetic waves. Investigations of Brewster metafilms as well as Huygens metasurfaces would yield further advances in controlling electromagnetic waves.

\begin{acknowledgments}
This research was supported by Grant for Basic Science Research Projects from The Sumitomo Foundation, and JSPS KAKENHI Grant Numbers JP18H03690, JP21K04192, and Japan Science and Technology Agency (CREST JPMJCR2101). 
\end{acknowledgments}


%

\end{document}